\documentclass{JHEP3}
\usepackage{mathrsfs}

\usepackage{amsfonts}
\usepackage{amsmath}
\usepackage{amssymb}
\usepackage{amsmath,amssymb,epsfig}
\numberwithin{equation}{section}
\usepackage{cite}

\newcommand{\be}{\begin{equation}}
\newcommand{\bea}{\begin{eqnarray}}
\newcommand{\eea}{\end{eqnarray}}
\newcommand{\ba}{\begin{array}}
\newcommand{\ea}{\end{array}}
\newcommand{\ee}{\end{equation}}

\allowdisplaybreaks[4]
\title{Relations Between Closed String Amplitudes at Higher-order Tree Level and Open String Amplitudes}

\author{Yi-Xin Chen, Yi-Jian Du and Qian Ma \\
Zhejiang Institute of Modern Physics, Zhejiang University\\
Hangzhou 310027, P. R. China\\
E-mail:\email{yxchen@zimp.zju.edu.cn},
\email{yjdu@zju.edu.cn},\email{mathons@gmail.com} }

 \abstract{KLT relations almost factorize closed string amplitudes on $S_2$ by two open string tree amplitudes which correspond to
 the left- and the right- moving sectors. In this paper, we
 investigate string amplitudes on $D_2$ and $RP_2$. We find that KLT
 factorization relations do not hold in these two cases. The relations between
 closed and open string amplitudes have new forms. On $D_2$ and $RP_2$, the
 left- and the right- moving sectors are connected into a single
 sector. Then an amplitude with closed strings on $D_2$ or $RP_2$ can be given by one open
 string tree amplitude except for a phase factor. The
 relations depends on the topologies of the world-sheets.
Under T-duality, the relations on $D_2$ and $RP_2$ give the
amplitudes between closed strings scattering from D-brane and
O-plane respectively by open string partial amplitudes.
  In the low energy limits of these two cases, the factorization
relations for
 graviton amplitudes do not hold. The amplitudes for gravitons must
 be given by the new relations instead.
 }
\begin{document}

\section{Introduction}
Superstring theories are theories without ultraviolet divergences.
They contain both gravitational and gauge interactions as low
energy limits\cite{1, 2}. Thus they offer a possible solution to
the problem of unifying all of the fundamental interactions in a
consistent quantum theory. In string theory, gravitons are
massless states of closed strings and gauge particles are massless
states of open strings. To study the relations between gravity and
gauge field, we should explore the relations between closed and
open strings. The duality between open and closed strings\cite{3,
4, 5, 6, 7, 8} also motivates us to explore the relations between
closed and open strings.

\begin{figure}[tbp]
\begin{center}
\includegraphics[width=0.7\textwidth]{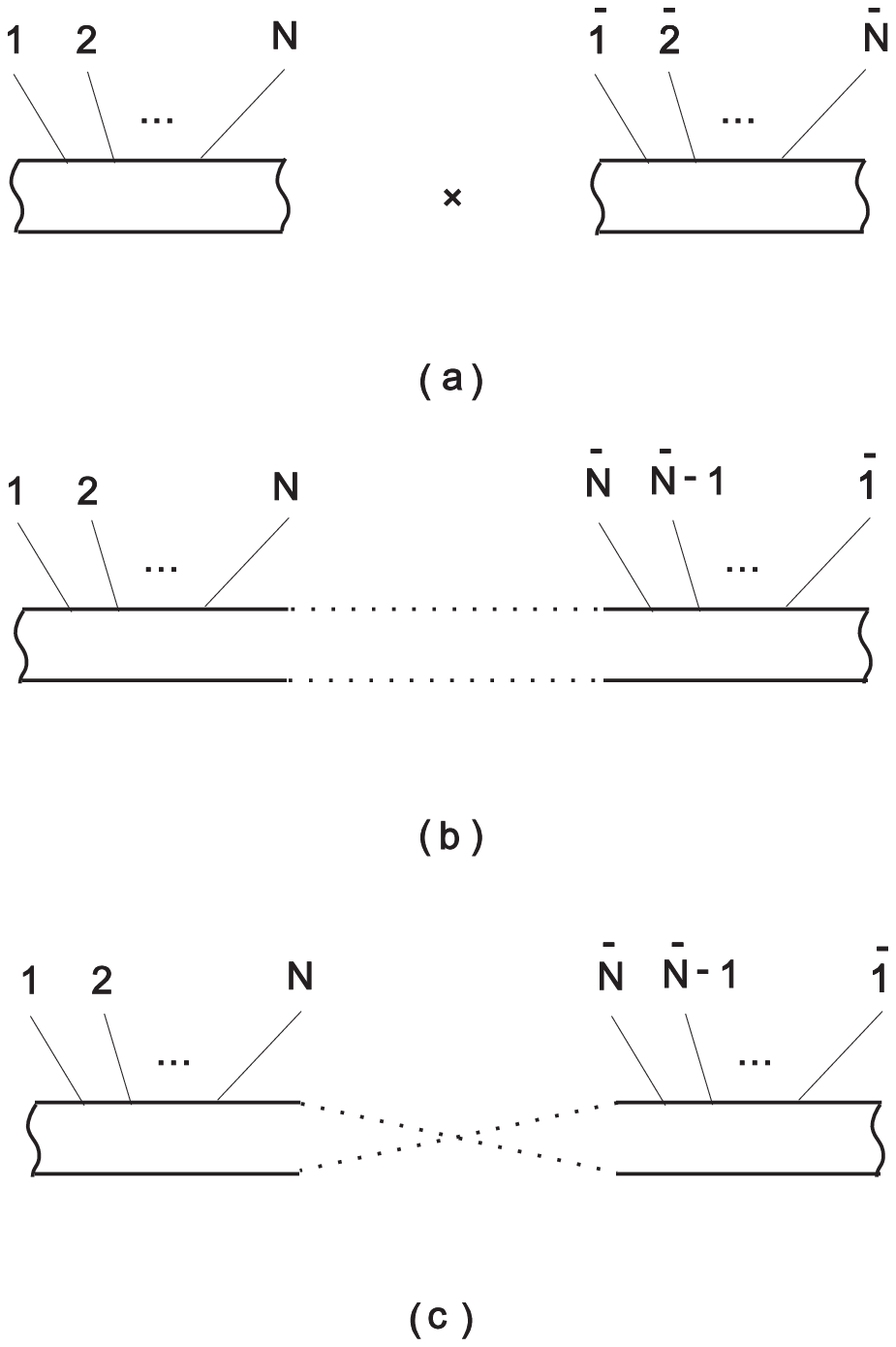}
\end{center}
\caption{(a) A closed string amplitude on $S_2$ can be factorize by
two open string tree amplitudes corresponding to the left- and
right-sectors. (b) A closed string amplitude on $D_2$ can be given
by connecting the open string world-sheets for the two sectors with
a time reverse in the right-moving sector. (c) A closed string
amplitude on $RP_2$ can be given by connecting the open string
world-sheets for the two sectors with a time reverse and a twist in
the right-moving sector.} \label{fig1}
\end{figure}

The most simple relation is any excited mode of a free closed string
$\left|N_L, N_R\right>\otimes \left|p\right>$ can be factorized by
left- and right- moving open string excited modes:
\begin{equation}
\left|N_L\right>\otimes\left|N_R\right>\otimes\left|p\right>.
\end{equation}
However, when we consider the interactions among strings, there are
nontrivial relations between closed and open string amplitudes. The
first nontrivial relation was given by Kawai, Lewellen and
Tye\cite{9}. They express an amplitude for $N$ closed strings on
sphere($S_2$) by the following equation\footnote{We use
$\epsilon_{\alpha\beta}$ to denote all the polarization tensors for
convenience. $\alpha$ correspond to the left indices and $\beta$
correspond to the right indices. If there are open strings on the
boundary of $D_2$, we use $\epsilon_{\alpha\beta\gamma}$ to denote
all the polarization tensors for convinience. $\gamma$ correspond to
the indices of open strings.}:
\begin{equation}\label{KLT relations}
\mathscr{A}_{S_2}^{(N)}=\epsilon_{\alpha\beta}\mathscr{A}_{S_2}^{(N)\alpha\beta}=\left(\frac{i}{2}\right)^{N-3}\kappa^{N-2}\epsilon_{\alpha\beta}
\sum\limits_{P,
P'}\mathscr{M}^{(N)\alpha}(P)\mathscr{\bar{M}}^{(N)\beta}(P')e^{i\pi
F(P, P')}.
\end{equation}
Here $\mathscr{A}_{S_2}^{N}$ is the amplitude for $N$ closed
strings on $S_2$ and $\mathscr{A}_{S_2}^{(N)\alpha\beta}$ is the
closed string amplitude without polarization tensors.
$\mathscr{M}^{(N)\alpha}(P)$ and
$\mathscr{\bar{M}}^{(N)\beta}(P')$ are the open string partial
amplitudes on $D_2$ corresponding to the left- and right-moving
sectors respectively. They are dependent on the orderings of the
external legs. If we sum over the orderings $P$ and $P'$, we get
the total amplitudes $\sum\limits_P\mathscr{M}^{(N)\alpha}(P)$ and
$\sum\limits_{P'}\mathscr{\bar{M}}^{(N)\beta}(P')$ for the left-
and the right-moving open strings respectively. Then we can see,
except for a phase factor, a closed string amplitude on $S_2$ can
be factorized by two open string tree amplitudes corresponding to
the left- and right-moving sectors(see fig. \ref{fig1}.(a)).
There is no interaction between left- and right-moving open
strings. Any closed string polarization tensor has left and right
indices, they correspond to the left- and the right-moving modes
respectively. The left and right indices of polarization tensors
must contract with the indices in the amplitude for left- and
right-moving open strings respectively. The phase factor is
entirely independent of which open and closed string theories we
are considering. It only depends on $P$ and $P'$. Contour
deformations can be used to reduce the number of the terms in
eq.\eqref{KLT relations}. The number of the terms can be reduced
to
\begin{equation}\label{KLT term reduction1}
(N-3)!(\frac{1}{2}(N-3))!(\frac{1}{2}(N-3))!, \text{$N$ odd},
\end{equation}
and
\begin{equation}\label{KLT term reduction2}
(N-3)!(\frac{1}{2}(N-2))!(\frac{1}{2}(N-4))!, \text{$N$ even}.
\end{equation}
\newline
\newline

In the low energy limits, the massive modes decouple. Only
massless states are left. Then KLT relations can be used to
factorize the amplitudes for gravitons into products of two
amplitudes for gauge particles. Gauge theory has a better
ultraviolet behavior than gravity. Then KLT relations can be used
to investigate the ultraviolet properties of gravity. Researches
with KLT relations support that $N=8$ supergravity may be
finite\cite{11, 12, 13, 14, 15, 16}. However, a question arises:
Do KLT factorization relations hold for any gravity amplitude? In
string theory, to calculate the S-matrix, we should sum over all
the topologies of world-sheets. $S_2$ is just the most simple
topology. If we consider other topologies, we should reconsider
the relations between closed and open strings. Then the question
becomes: Do the factorization relations hold for any topology?

Earlier works\cite{17, 25, 26, 27} have given some insights into the
relations on Disk($D_2$). In \cite{17}, some examples of the
relations on $D_2$ are given. In \cite{25, 26, 27}, the most simple
process of D-brane and closed string interactions are discussed. In
the paper by Garousi and Myers\cite{25}, they found that the
two-point scattering amplitudes of closed strings from a D-brane in
Type II theory is identical with the four-point open string
amplitudes upon a certain identification between the momenta and
polarizations. In \cite{26, 27}, The amplitude for one closed string
and two open strings attached to a D-brane are calculated. They
shown that this amplitude are also identical with the four-point
open string amplitude. In these examples, The KLT factorization
relations do not hold.  Then they imply the KLT factorization
relations may not hold for general amplitudes on $D_2$. The
amplitudes on real projective plane($RP_2$) have similar structures
with open string amplitudes\cite{30}. In fact, Both $D_2$ and $RP_2$
can be obtained by a sphere with a $\mathbb{Z}_2$ identification.
Then the KLT factorization relations may also not hold in the $RP_2$
case.

In this paper, we consider the general amplitudes on $D_2$ and
$RP_2$. These two cases contribute to the higher-order tree
amplitude\cite{1} for closed strings. We find that the factorization
relations\eqref{KLT relations} do not hold on $D_2$ and $RP_2$. The
amplitudes with closed strings on $D_2$ and $RP_2$ can not be
factorized by the left- and the right-moving open string amplitudes.
The amplitudes satisfy new relations. Particularly, an amplitude for
$N$ closed strings on $D_2$ can be given by an amplitude for $2N$
open strings:
\begin{equation}\label{A_(D_2)^(N,0)}
\mathscr{A}_{D_2}^{(N)}=\epsilon_{\alpha\beta}\mathscr{A}_{D_2}^{(N)\alpha\beta}=\left(\frac{i}{4}\right)^{N-1}\kappa^{N-1}\epsilon_{\alpha\beta}\sum\limits_P\mathscr{M}^{(2N)\alpha\beta}(P)e^{i\pi\Theta(P)}.
\end{equation}
In this equation, $\mathscr{M}^{(2N)\alpha\beta}(P)$ is the tree
amplitude for $2N$ open strings. $N$ open strings come from the
left-moving sector and the other $N$ open strings come from the
right-moving sector. The left- and the right-moving sectors are not
independent of each other. The two sectors are connected into a
single sector. Then the left indices contract with the right
indices. The reason is that the left-(right-)moving waves must be
reflected at the boundary of $D_2$ and then become
right-(left-)moving waves. Then the interactions between the
left-(right-)moving waves and their reflected waves become
interactions between the two sectors. If there are open strings on
the boundary of $D_2$, the left- and the right-moving sectors of
closed strings also interact with the open strings, then an
amplitude for $N$ closed strings and $M$ open strings on $D_2$ can
be given by a tree amplitude for $2N+M$ open strings except for a
phase factor:
\begin{equation}
\label{A_(D_2)^(N,M)} \mathscr{A}_{D_2}^{(N,
M)}=\epsilon_{\alpha\beta\gamma}\mathscr{A}_{D_2}^{(N,
M)\alpha\beta\gamma}=\left(\frac{i}{4}\right)^{N-1}\kappa^{N-1}g^M\epsilon_{\alpha\beta\gamma}
\sum\limits_P\mathscr{M}^{(2N,M)\alpha\beta\gamma}(P)e^{i\pi\Theta'(P)}.
\end{equation}

The amplitudes on $RP_2$ can also be factorized by one amplitude for
open strings:
\begin{equation}\label{A_RP2^(N)}
\mathscr{A}_{RP_2}^{(N)}=\epsilon_{\alpha\beta}\mathscr{A}_{RP_2}^{(N)\alpha\beta}=-\left(\frac{i}{4}\right)^{N-1}\kappa^{N-1}\epsilon_{\alpha\beta}\sum\limits_P\mathscr{M}^{(2N)\alpha\beta}(P)e^{i\pi\Theta(P)}.
\end{equation}
In this case, there is a crosscap but not a boundary here. However,
the left-(right-) moving waves are also reflected at the crosscap
and turn into the right-(left-)moving waves. Then there are also
interactions between left- and right-moving sectors of closed
strings. The two sectors are connected into one single sector again.
The phase factors in \eqref{A_(D_2)^(N,M)} and \eqref{A_RP2^(N)} are
complicated in concrete calculations. By considering the contour
deformations, the number of the terms can be reduced\cite{9, 29}. It
is noticed that the relations on $D_2$  are same with on $RP_2$
except for a minus. In a theory containing both $D_2$ and $RP_2$,
the two amplitudes cancel out. Under a T-duality, the relation
\eqref{A_(D_2)^(N,M)} gives the amplitude for $N$ closed strings and
$M$ open strings attached to a D-brane by pure open string
amplitudes while the relation\eqref{A_RP2^(N)} gives the amplitude
for $N$ closed strings scattering from an O-plane by pure open
string amplitudes. In this case, the amplitudes on $D_2$ and $RP_2$
can not cancel out.

An important fact will be used in our paper is that the amplitudes
with closed strings are invariant under conformal transformations in
each single sector. This allows us to transform the form of the
interactions between left- and right-moving sectors. After some
appropriate transformation in one sector, the interactions between
left- and the right-moving sectors have the same form with
interactions between open strings in a same sector. Then we can
treat the two sectors of $N$ closed strings as a single sector with
$2N$ open strings.

In the low energy limit of an unoriented open string theory, the
amplitudes for $N$ closed strings on $D_2$, $RP_2$ and $S_2$
contribute to the tree amplitudes for $N$ gravitons. In this case,
we can not only use KLT factorization relations on $S_2$ but also
use the relations on $D_2$ and $RP_2$ to calculate the tree
amplitudes for gravitons. The amplitudes for $N$ closed strings and
$M$ open strings on $D_2$ become tree amplitudes for $N$ gravitons
and $M$ gauge particles. Then the gauge-gravity interactions can be
given by pure gauge interactions.

 The structure of this paper is as follows. In section \ref{relations
on D_2} we will consider the correlation functions and the
amplitudes on $D_2$. We will show  KLT factorization relations do
not hold on $D_2$. We will give the relations between closed string
amplitudes on $D_2$ and open string tree amplitudes. We will also
give the relations in the case of $N$ closed strings and $M$ open
strings inserted on $D_2$. In section \ref{relations on RP_2} we
will consider $RP_2$. We will show KLT factorization relations do
not hold on $RP_2$. The relations between amplitudes on $RP_2$ and
open string amplitudes will be given in this section. Our conclusion
will be given in section\ref{conclusion}.
\section{Relations between amplitudes on $D_2$ and open
string tree amplitudes} \label{relations on D_2} In this section, we
will show the correlation functions on $D_2$ can not be factorized
into the left- and the right-moving sectors. The two sectors are
connected together. Then we will give the relations between
amplitudes on $D_2$\cite{1, 2, 18, 19, 22} and open string tree
amplitudes\cite{1, 2, 22, 23, 24, 25, 26, 27, 28}.

In string theory, vertex operator for any closed string can be
given as
\begin{equation}
\mathscr{V}(\omega,
\bar{\omega})=\mathscr{V}_L(\omega)\tilde{\mathscr{V}}_R(\bar{\omega})\mathscr{V}_0(\omega,
\bar{\omega}),
\end{equation}
where $\omega=\tau+i\sigma$. $\mathscr{V}_L$ and $\mathscr{V}_R$
are nonzero modes of open string vertex operators. They correspond
to the left- and the right-moving sectors. $\mathscr{V}_0(\omega,
\bar{\omega})$ correspond to the zero modes. Thus, the closed
string vertex operators can be factorized by two open string
vertex operators corresponding to the left- and the right-moving
sectors(except for the zero modes).
\newline
\begin{figure}[tbp]
\begin{center}
\includegraphics[width=0.5\textwidth]{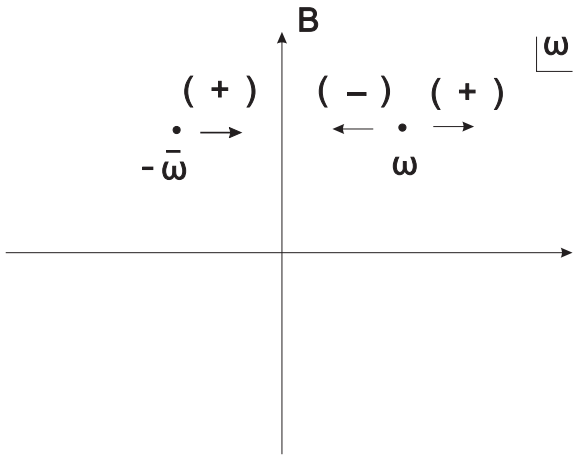}
\end{center}
\caption{Only the annihilation modes are reflected at the boundary
of $D_2$.} \label{fig2}
\end{figure}
\newline
Now we consider the correlation function of vertex operators. On
$S_2$ the left- and the right-moving waves are independent of each
other. Then a correlation function on $S_2$ can be factorized by
the left- and the right-moving sectors\cite{1, 2}. However, when
we add a boundary to $S_2$, we get $D_2$. The left- and the
right-moving waves must be reflected at the boundary of $D_2$. The
reflection waves of the left-moving waves are in the right-moving
sector and the reflection waves of the right-moving waves are in
the left-moving sector. Waves must interact with their reflection
waves, then their must be interactions between the two sectors. To
see this, we should use the boundary state\cite{20, 21} to give
the correlation functions on $D_2$. The correlation function for N
closed strings on $D_2$ is
\begin{equation}\label{D2 correlation}
\left<0\mid\mathscr V_N(\omega,
\tilde{\omega})...\mathscr{V}_1(\omega, \tilde{\omega})\mid
B\right>,
\end{equation}
where  $\mid B\rangle\equiv B\mid 0\rangle$ is the boundary state
for $D_2$. In this paper, for convenience, we use the bosonized
vertex operators\footnote{Here $\phi_i(z) (i=1...5)$ and
$\tilde{\phi}_i(\bar{z}) (i=1...5)$ are bosonic fields. They are
used to bosonize holomorphic and antiholomorphic fermionic fields
and spinor fields. $\phi_6(z)$ and $\tilde{\phi}_6(\bar{z})$ are
used to bosonize the holomorphic and antiholomorphic
superconformal ghost respectively. $\epsilon$ and $\bar{\epsilon}$
correspond to the components of polarization tensors contracting
with bosonic fields $\partial X$  and $\bar{\partial}X$
respectively. $\varepsilon$ and $\bar{\varepsilon}$ correspond to
the components contracting with $\partial\phi$ and
$\bar{\partial}\tilde{\phi}$ respectively. We pick up the pieces
multilinear in $\epsilon$, $\bar{\epsilon}$ and $\varepsilon$,
$\bar{\varepsilon}$, then replace these polarization vectors by
the polarization tensor of the vertex operator.
 $\lambda_i'=i\lambda_i$ and
$\tilde{\lambda}_i'=i\tilde{\lambda}_i$ (i=1...5) are vectors in the
weight lattice\cite{18, 19} of the left- and right-moving sectors
respectively. $q$ and $\tilde{q}$ are the $\gamma$ ghost number in
the left- and right-moving sectors respectively. We use $\circ$ to
denote the inner product in the five dimensional weight space and
use $\cdot$ to denote the inner product in the space-time.
\\Physical vertices containing higher derivatives can be transformed
into the vertices with only first derivatives. In fact we can do
partial integrals to reduce the order of the derivatives. After the
integrals on the world-sheet, the surface terms turn to zero.
Redefine the polarization tensor, the vertices then turn to those
only contain first derivatives. }
\begin{equation}\label{bosonized vertex operator}
\begin{split}
\mathscr{V}(\omega,\bar{\omega})=&:\exp{(q\phi_6+\tilde{q}\tilde{\phi}_6)}
\\&\exp{(i\lambda\circ\phi+i\sum\limits_{i=1}^{m}\varepsilon^i\circ\partial\phi_i
+i\tilde{\lambda}\circ\tilde{\phi}+i\sum\limits_{i=1}^{\tilde{m}}\bar{\varepsilon}^i\circ\bar{\partial}\tilde{\phi_i}
)}
\\&\exp{(ik\cdot
X+i\sum\limits_{i=1}^n\epsilon^i\cdot\partial
X+i\sum\limits_{j=1}^{\tilde{n}}\bar{\epsilon}^j\cdot\bar{\partial}X)}(\omega,\bar{\omega}):|_{multilinear}
\end{split}
\end{equation}
With the definition of normal ordering, we have
\begin{equation}
\mathscr{V}(\omega,\bar{\omega})=\mathscr{V}^{(+)}_L(\omega)\mathscr{V}^{(-)}_L(\omega)\tilde{\mathscr{V}}^{(+)}_L(\bar{\omega})\tilde{\mathscr{V}}^{(-)}_R(\bar{\omega})\mathscr{V}_0(\omega,
\bar{\omega}),
\end{equation}
where $(+)$ and $(-)$ correspond to the creation modes and the
annihilation modes respectively. In $\mathscr{V}_0$, we consider
$x$ as creation operator and $p$ as annihilation operator. Then in
the normal ordering, $x$ must on the left of $p$.
  The
bosonized boundary operator is\footnote{Here, we only consider
Neumann boundary condition for convenience. The case with Dirichlet
boundary conditions has similar relations.}
\begin{equation}
 \begin{split}
B&=\exp{(\sum\limits_{n=1}^\infty
a_n^{\dagger}\cdot\tilde{a}_{n}^{\dagger})}\otimes\exp{(\sum\limits_{n=1}^\infty
b_n^{\dagger}\circ\tilde{b}_{n}^{\dagger})}\otimes\exp{(\sum\limits_{n=1}^\infty
c_n^{\dagger}\tilde{c}_{n}^{\dagger})},
\end{split}
\end{equation}
where $a^{\dag}$ and $\tilde{a}^{\dag}$ are creation modes of $X$,
$b^{\dag}$ and $\tilde{b}^{\dag}$ are creation modes of $\phi_i$
and $\tilde{\phi_i}$ respectively,  $c^{\dag}$ and
$\tilde{c}^{\dag}$ are creation modes of $\phi_6$ and
$\tilde{\phi}_6$ respectively. To get the correlation function on
$D_2$ we substitute the bosonized vertex operators and the
bosonized boundary operators into \eqref{D2 correlation}.
 We can move the boundary operator $B$ to the left of all the
vertex operators. Then use the creation operators in $B$ to
annihilate the state $\langle 0 \mid$. Because $B$ is constructed
by creation operators, it commutes with the creation modes and the
zero modes of the vertex operators and does not commute with the
annihilation modes of the vertex operators. It means only  the
annihilation modes are reflected at the boundary(see fig.
\ref{fig2}). When we move $B$ to the left of the annihilation
modes of the vertex operators $\mathscr{V}^{(-)}_L(\omega)$ and
$\tilde{\mathscr{V}}^{(-)}_R(\bar{\omega})$, the ''images'' of the
annihilation modes $\tilde{\mathscr{V}}^{(+)}_L(-\omega)$ and
$\mathscr{V}^{(+)}_R(-\bar{\omega})$ are created respectively.
Though $\tilde{\mathscr{V}}^{(+)}_L(-\omega)$ is depend on
$\omega$, it is constructed by $\tilde{a}^{\dag}$,
$\tilde{b}^{\dag}$ and $\tilde{c}^{\dag}$. It must interact with
operators constructed by $\tilde{a}$, $\tilde{b}$ and $\tilde{c}$.
In a similar way, $\mathscr{V}^{(+)}_R(-\bar{\omega})$ must
interact with operators constructed by $a$, $b$ and $c$. Then the
correlation function can be factorized as
\begin{equation}\label{correlation function on D2}
\begin{split}
&\left<\mathscr{V}^{(+)}_L(\omega_N)\mathscr{V}^{(-)}_L(\omega_N)\mathscr{V}^{(+)}_R(-\bar{\omega}_N)...\mathscr{V}^{(+)}_L(\omega_1)\mathscr{V}^{(-)}_L(\omega_1)\mathscr{V}^{(+)}_R(-\bar{\omega}_1)\right>
\\\times &\left<\tilde{\mathscr{V}}^{(+)}_R(\bar{\omega}_N)\tilde{\mathscr{V}}^{(-)}_R(\bar{\omega}_N)\tilde{\mathscr{V}}^{(+)}_L(-\omega_N)...\tilde{\mathscr{V}}^{(+)}_R(\bar{\omega}_1)\tilde{\mathscr{V}}^{(-)}_R(\bar{\omega}_1)\tilde{\mathscr{V}}^{(+)}_L(-\omega_1)\right>
\\\times &\left<\mathscr{V}_0(\omega_N, \tilde{\omega}_N)...\mathscr{V}_0(\omega_1,
\tilde{\omega}_1)\right>.
\end{split}
\end{equation}
Here, the first correlation function only contain operators
constructed by $a$, $b$, $c$ and $a^{\dag}$, $b^{\dag}$,
$c^{\dag}$, the second correlation function only contain operators
constructed by $\tilde{a}$, $\tilde{b}$, $\tilde{c}$ and
$\tilde{a}^{\dag}$, $\tilde{b}^{\dag}$, $\tilde{c}^{\dag}$, the
third correlation function only contain operators constructed by
zero modes. Though the nonzero modes are factorized into two
correlation functions, both of them contain the interactions
between left- and the right-moving sectors. Actually, in the first
correlation function, if we move $\mathscr{V}^{(-)}_L(\omega_i)$
to the right of $\mathscr{V}^{(+)}_L(\omega_j)$, we get the
interaction in the left-moving sector and if we move
$\mathscr{V}^{(-)}_L(\omega_i)$ to the right of
$\mathscr{V}^{(+)}_R(-\bar{\omega}_j)$, we get the interactions
between the two sectors. In the same way, the second correlation
function contain interactions in the right-moving sector and the
interactions between left- and right-moving sectors. The
interactions between the two sectors are just the interactions
between vertex operators and their images. Then the correlation
function on $D_2$ can not be factorized by the the two sectors,
interactions between the two sectors connect them together.

To get the relations between amplitudes, we should calculate the
correlation function, then integral over the fundamental region
and divide the integrals by the volume of conformal Killing
group\cite{1, 2, 22, 23}. for convenience, we use the $z$
coordinate instead of $\omega$ coordinate. They are connected by a
conformal transfermation:
\begin{equation}
z=e^{\omega}.
\end{equation}
Then the amplitude for $N$ closed strings on $D_2$ becomes
\begin{align}\label{AD2superstring}
\nonumber\mathscr{A}_{D_2}^{(N,0)}&=\kappa^{N-1}\int_{|z|<1}
\prod\limits_{i=1}^Nd^2z_i\frac{|1-z_o\bar{z_o}|^2}{2\pi d^2z_o}
\\&\nonumber\times\prod\limits_{s>r}(z_s-z_r)^{\frac{\alpha'}{2}k_r\cdot
k_s+\lambda_r\circ\lambda_s-q_rq_s}(\bar{z}_r-\bar{z}_s)^{\frac{\alpha'}{2}k_r\cdot
k_s+\tilde{\lambda}_r\circ\tilde{\lambda}_s-\tilde{q}_r\tilde{q}_s}
\prod\limits_{r,
s}(1-(z_r\bar{z}_s)^{-1})^{\frac{\alpha'}{2}k_r\cdot
k_s+\lambda_r\circ\tilde{\lambda}_s-q_r\tilde{q}_s}
\\&\nonumber\times \exp{\sum\limits_{r=1}^N\left(\sum\limits_{i=1}^{n_r}\sum\limits_{j=1}^{\tilde{n}_s}\left(-\frac{\alpha'}{2}\right)\epsilon_r^{(i)}\cdot\bar{\epsilon}_r^{(j)}
-\sum\limits_{i=1}^{m_r}\sum\limits_{j=1}^{\tilde{m}_s}\varepsilon_r^{(i)}\circ\bar{\varepsilon}_r^{(j)}\right)(1-|z_r|^2)^{-2}}
\\&\times \exp{\sum\limits_{s>r}\left[\left(\sum\limits_{i=1}^{\tilde{n}_r}\sum\limits_{j=1}^{n_s}\left(-\frac{\alpha'}{2}\right)\bar{\epsilon}_r^{(i)}\cdot\epsilon_s^{(j)}
-\sum\limits_{i=1}^{\tilde{m}_r}\sum\limits_{j=1}^{m_s}\bar{\varepsilon}_r^{(i)}\circ\varepsilon_s^{(j)}\right)(1-\bar{z}_rz_s)^{-2}+c.c.\right]}
\\&\nonumber\times\exp{\left[-\sum\limits_{s>r}\left(\sum\limits_{i=1}^{n_r}\sum\limits_{j=1}^{n_s}\left(-\frac{\alpha'}{2}\right)\epsilon_r^{(i)}\cdot\epsilon_s^{(j)}-\sum\limits_{i=1}^{m_r}\sum\limits_{j=1}^{m_s}\varepsilon_r^{(i)}\circ\varepsilon_s^{(j)}\right)(z_s-z_r)^{-2}+c.c.
\right]}
\\&\nonumber\times \exp{\sum\limits_{r\neq s}\left[\left(\sum\limits_{i=1}^{n_s}\left(-\frac{\alpha'}{2}\right)k_r\cdot\epsilon_s^{(i)}
-\sum\limits_{i=1}^{m_s}\lambda_r\circ\varepsilon_s^{(i)}\right)((z_r-z_s)^{-1}+(\bar{z_r}^{-1}-z_s)^{-1})+c.c.\right]}
\\&\nonumber\times
\exp{\sum\limits_{r=1}^N\left[\left(\left(-\frac{\alpha'}{2}\right)k_r\cdot\sum\limits_{i=1}^{n_r}\epsilon_r^{(i)}
-\lambda_r\circ\sum\limits_{i=1}^{m_r}\varepsilon_r^{(i)}\right)((\bar{z_r}^{-1}-z_r)^{-1}+{z_r}^{-1})+c.c.\right]}|_{multilinear},
\end{align}
where we have
$\sum\limits_{r=1}^N\lambda_r=\sum\limits_{r=1}^N\tilde{\lambda}_r=0$,
$\sum\limits_{r=1}^Nk_r=0$ and
$\sum\limits_{r=1}^N(q_r+\tilde{q}_r)=-2$ correspond to the
conservation of fermion number, the conservation of momentum and the
fact that background superghost number is $-2$. $\frac{2\pi
d^2z_o}{|1-z_o\bar{z_o}|^2}$ is the volume element of the
CKG\footnote{To divide the amplitude by the volume of CKG, we can
fix three real coordinate. We can also fix two real coordinate or
one complex coordinate, then divide the amplitude by volume of the
one-parameter subgroup left. The two method are equivalence. Here,
we use the second method to fix $z_1=z_o$.}, it can be used to fix
one complex coordinate.

An integral over the fundamental region $|z|<1$ is equal to an
integral over the other fundamental region $|z|>1$. So we can use
${\left(\frac{1}{2}\right)}^{N-1}\int\limits_{\mathbb{C}}\prod\limits^N_{i=1}d^2z_i$
instead of the integrals over the unit disk. For any
$z_r=x_r+iy_r$, the $z_r$ integral can be given by
$\int\limits_{-\infty}^{\infty}dx_r\int\limits_{-\infty}^{\infty}dy_r$.
 We then follow the same steps as in\cite{9}. We rotate the contour of the $y$ integrals along the real axis to pure imaginary axis.  The fixed
point should be transformed simultaneously to guarantee the
conformal invariance. Define the new variables:
\begin{equation}\label{variable redefine}
\begin{split}
\xi_1=\xi_o=x_o+iy_o &, \eta_1=\eta_o=x_o-iy_o, \\\xi_r\equiv
x_r+iy_r &, \eta_r\equiv x_r-iy_r\text{   }(r>1).
\end{split}
\end{equation}
Then the integrals become real integrals:
\begin{align}\label{AD2real integrals}
\nonumber\mathscr{A}_{D_2}^{(N,0)}&=\kappa^{N-1}\left(\frac{1}{2}\right)^{N-1}\int
\prod\limits_{i=1}^Nd\xi_id\eta_i\frac{|1-\xi_o\eta_o|^2}{2\pi
d\xi_od\eta_o}
\\&\nonumber\times\prod\limits_{s>r}(\xi_s-\xi_r)^{\frac{\alpha'}{2}k_r\cdot
k_s+\lambda_r\circ\lambda_s-q_rq_s}(\eta_r-\eta_s)^{\frac{\alpha'}{2}k_r\cdot
k_s+\tilde{\lambda}_r\circ\tilde{\lambda}_s-\tilde{q}_r\tilde{q}_s}
\prod\limits_{r,
s}(1-(\xi_r\eta_s)^{-1})^{\frac{\alpha'}{2}k_r\cdot
k_s+\lambda_r\circ\tilde{\lambda}_s-q_r\tilde{q}_s}
\\&\nonumber\times \exp{\sum\limits_{r=1}^N\left(\sum\limits_{i=1}^{n_r}\sum\limits_{j=1}^{\tilde{n}_s}\left(-\frac{\alpha'}{2}\right)\epsilon_r^{(i)}\cdot\bar{\epsilon}_r^{(j)}
-\sum\limits_{i=1}^{m_r}\sum\limits_{j=1}^{\tilde{m}_s}\varepsilon_r^{(i)}\circ\bar{\varepsilon}_r^{(j)}\right)(1-\xi_r\eta_r)^{-2}}
\\&\times \exp{\sum\limits_{s>r}\left[\left(\sum\limits_{i=1}^{\tilde{n}_r}\sum\limits_{j=1}^{n_s}\left(-\frac{\alpha'}{2}\right)\bar{\epsilon}_r^{(i)}\cdot\epsilon_s^{(j)}
-\sum\limits_{i=1}^{\tilde{m}_r}\sum\limits_{j=1}^{m_s}\bar{\varepsilon}_r^{(i)}\circ\varepsilon_s^{(j)}\right)(1-\eta_r\xi_s)^{-2}+c.c.\right]}
\\&\nonumber\times\exp{\left[-\sum\limits_{s>r}\left(\sum\limits_{i=1}^{n_r}\sum\limits_{j=1}^{n_s}\left(-\frac{\alpha'}{2}\right)\epsilon_r^{(i)}\cdot\epsilon_s^{(j)}-\sum\limits_{i=1}^{m_r}\sum\limits_{j=1}^{m_s}\varepsilon_r^{(i)}\circ\varepsilon_s^{(j)}\right)(\xi_s-\xi_r)^{-2}+c.c.
\right]}
\\&\nonumber\times \exp{\sum\limits_{r\neq s}\left[\left(\sum\limits_{i=1}^{n_s}\left(-\frac{\alpha'}{2}\right)k_r\cdot\epsilon_s^{(i)}
-\sum\limits_{i=1}^{m_s}\lambda_r\circ\varepsilon_s^{(i)}\right)((\xi_r-\xi_s)^{-1}+(\eta_r^{-1}-\xi_s)^{-1})+c.c.\right]}
\\&\nonumber\times
\exp{\sum\limits_{r=1}^N\left[\left(\left(-\frac{\alpha'}{2}\right)k_r\cdot\sum\limits_{i=1}^{n_r}\epsilon_r^{(i)}
-\lambda_r\circ\sum\limits_{i=1}^{m_r}\varepsilon_r^{(i)}\right)((\eta_r^{-1}-\xi_r)^{-1}+{\xi_r}^{-1})+c.c.\right]}|_{multilinear},
\end{align} The real variables $\xi_r$ correspond to the
left-moving sector and $\eta_r$ correspond to the left-moving
sector.
 An open string tree amplitude for $M$ bosonized vertices has the
form
\begin{equation}\label{open string amplitude}
\begin{split}
\mathscr{M}_{D_2}^{(N)}&=(g)^{M-2}\int \prod\limits_{i=1}^{M}d
x_i\frac{|x_a-x_b||x_b-x_c||x_c-x_a|}{d x_ad x_bd
x_c}\prod\limits_{s>r}|x_s-x_r|^{2\alpha'k_r\cdot
k_s}(x_s-x_r)^{\lambda_r\circ\lambda_s-q_rq_s}
\\&\times\exp{\left[\sum\limits_{s>r}\left(\sum\limits_{i=1}^{n_r}\sum\limits_{j=1}^{n_s}\left(2\alpha'\right)\epsilon_r^{(i)}\cdot\epsilon_s^{(j)}+\sum\limits_{i=1}^{m_r}\sum\limits_{j=1}^{m_s}\varepsilon_r^{(i)}\circ\varepsilon_s^{(j)}\right)(x_s-x_r)^{-2}
\right]}
\\&\times \exp{\left[\sum\limits_{r\neq s}\left(\sum\limits_{i=1}^{n_s}\left(-2\alpha'\right)k_r\cdot\epsilon_s^{(i)}
-\sum\limits_{i=1}^{m_s}\lambda_r\circ\varepsilon_s^{(i)}\right)(x_r-x_s)^{-1}\right]}|_{multilinear},
\end{split}
\end{equation}
where $g$ is the coupling constant for open strings, it can be
related with closed string coupling constant by $\kappa\sim g^2$.
 By comparing \eqref{AD2real integrals} with the open string amplitude\eqref{open string amplitude}, we can see, the interactions in one sector
can be considered as interactions between open strings. The
interactions between left- and right-moving sectors look like
those between open strings inserted at $\xi_r$ and
$(\eta_s)^{-1}$. $\eta_s$ can be considered as the coordinates of
the right-moving open string. Then in the $(\tau, \sigma)$
coordinate, ${\eta_s}^{-1}=e^{-\tau}$ can be considered as a time
reverse in the right-moving sector. Thus the interactions between
the two sectors can be regarded as interactions between left- and
right-moving open strings with a time reverse in the right moving
sector(see fig. \ref{fig1}.(b)). The amplitude\eqref{AD2real
integrals} then can be considered as an amplitude for $2N$ open
strings. $N$ of them correspond to the left-moving sector and the
other $N$ of them correspond to the right-moving sector. In the
amplitude we have a time reverse in the right-moving sector.

From fig. \ref{fig1}.(b), we can see, if we reverse  the time in
the right-moving sector, we will get an open string tree
amplitude. In fact, we can replace all the ${\eta_r}^{-1}$ by
$\eta_r$.  By using the mass-shell condition\cite{18, 19} which is
determined by the conformal invariance in one sector, the
interactions between the two sectors as well as the interactions
in one sector become those between open strings.
 Define
\begin{equation}\label{variable redefine}
\xi_{r+N}\equiv\eta_r, k_{r+N}\equiv k_r,
\tilde{\lambda}_{r+N}\equiv\lambda_r,
\bar{\epsilon}_{r+N}\equiv\epsilon_r,
\bar{\varepsilon}_{r+N}\equiv\varepsilon_r.
\end{equation}
After the simultaneous transformations, the volume of CKG becomes
$\frac{1}{2\pi}\int\frac{d\xi_od\eta_o}{|\xi_o-\eta_o|^2}$. The
fixed points become $\xi_1=\xi_o$ and $\xi_{1+N}=\xi_o$. The
conformal Killing volume has another form
$\int\frac{dx_adx_bdx_c}{|x_a-x_b||x_b-x_c||x_c-x_a|}$, it can be
used to fix three real variables. We reset the fixed points at:
\begin{equation}
\xi_1=x_a=0, \xi_2=x_b=1, \xi_{2N}=x_c=\infty.
\end{equation}
The amplitude for $N$ closed strings on $D_2$ then becomes
\begin{equation}\label{AD_2N1without phase factor}
\begin{split}
\mathscr{A}_{D_2}^{(N,0)}&=\kappa^{N-1}{\left(\frac{i}{4}\right)}^{N-1}\int
\prod\limits_{i=1}^{2N}d\xi_i\frac{|\xi_a-\xi_b||\xi_b-\xi_c||\xi_c-\xi_a|}{d\xi_ad\xi_bd\xi_c}\prod\limits_{s>r}(\xi_s-\xi_r)^{\frac{\alpha'}{2}k_r\cdot
k_s}(\xi_s-\xi_r)^{\lambda_r\circ\lambda_s-q_rq_s}
\\&\times\exp{\left[\sum\limits_{s>r}\left(\sum\limits_{i=1}^{n_r}\sum\limits_{j=1}^{n_s}\left(2\alpha'\right)\epsilon_r^{(i)}\cdot\epsilon_s^{(j)}+\sum\limits_{i=1}^{m_r}\sum\limits_{j=1}^{m_s}\varepsilon_r^{(i)}\circ\varepsilon_s^{(j)}\right)(\xi_s-\xi_r)^{-2}
\right]}
\\&\times \exp{\left[\sum\limits_{r\neq s}\left(\sum\limits_{i=1}^{n_s}\left(-2\alpha'\right)k_r\cdot\epsilon_s^{(i)}
-\sum\limits_{i=1}^{m_s}\lambda_r\circ\varepsilon_s^{(i)}\right)(\xi_r-\xi_s)^{-1}\right]}|_{multilinear}e^{i\pi\Theta(P)}
.
\end{split}
\end{equation}
After taking an appropriate phase factor out, we get
\begin{equation}\label{AD_2N1}
\begin{split}
\mathscr{A}_{D_2}^{(N,0)}&=\kappa^{N-1}{\left(\frac{i}{4}\right)}^{N-1}\int
\prod\limits_{i=1}^{2N}d\xi_i\frac{|\xi_a-\xi_b||\xi_b-\xi_c||\xi_c-\xi_a|}{d\xi_ad\xi_bd\xi_c}\prod\limits_{s>r}|\xi_s-\xi_r|^{\frac{\alpha'}{2}k_r\cdot
k_s}(\xi_s-\xi_r)^{\lambda_r\circ\lambda_s-q_rq_s}
\\&\times\exp{\left[\sum\limits_{s>r}\left(\sum\limits_{i=1}^{n_r}\sum\limits_{j=1}^{n_s}\left(2\alpha'\right)\epsilon_r^{(i)}\cdot\epsilon_s^{(j)}+\sum\limits_{i=1}^{m_r}\sum\limits_{j=1}^{m_s}\varepsilon_r^{(i)}\circ\varepsilon_s^{(j)}\right)(\xi_s-\xi_r)^{-2}
\right]}
\\&\times \exp{\left[\sum\limits_{r\neq s}\left(\sum\limits_{i=1}^{n_s}\left(-2\alpha'\right)k_r\cdot\epsilon_s^{(i)}
-\sum\limits_{i=1}^{m_s}\lambda_r\circ\varepsilon_s^{(i)}\right)(\xi_r-\xi_s)^{-1}\right]}|_{multilinear}e^{i\pi\Theta(P)}
,
\end{split}
\end{equation}
where we have absorbed a factor $\frac{1}{2}$ into each $\epsilon$.
$\Theta(P)$ is defined as
\begin{equation}
\Theta(P)=\sum\limits_{s>r}2\alpha'k'_s\cdot
k'_r\theta(\xi_s-\xi_r),
\end{equation}
where $k'^{\mu}_r=\frac{1}{2}k^{\mu}_r$ is the momentum of the open
string and
\begin{align}
\theta(\xi_s-\xi_r)=
  \biggl\{
  \begin{array}{l}
    0 ( \xi_s>\xi_r) \\
    1 ( \xi_s<\xi_r)  \\
  \end{array}.
\end{align}
  From\eqref{open string amplitude} and
\eqref{AD_2N1} we can see amplitudes for $N$ closed strings on $D_2$
can be given by one open string tree amplitude for $2N$ open strings
except for a phase factor. The phase factor is caused by taking
absolute number of $(\xi_s-\xi_r)$ in
$(\xi_s-\xi_r)^{\frac{\alpha'}{2}k_r\cdot k_s}$. It is used to
guarantee the integrals in the right branch cut. It only depend on
the the orderings of the open strings. For a certain order $P$, the
phase factor decouple from the integrals. So we can break the
integrals into pieces, take the multilinear terms in $\epsilon$,
$\bar{\epsilon}$, $\varepsilon$ and $\bar{\varepsilon}$, replace the
polarization vectors by the polarization tensors of closed strings.
Then we get the relation between closed string amplitudes and
partial amplitudes for open strings on $D_2$:
\begin{equation}\label{D2relations}
\mathscr{A}_{D_2}^{(N,0)}=\kappa^{N-1}\epsilon_{\alpha\beta}\mathscr{A}_{D_2}^{(N)\alpha\beta}=\left(\frac{i}{4}\right)^{N-1}\kappa^{N-1}\epsilon_{\alpha\beta}\sum\limits_P\mathscr{M}^{(2N)\alpha\beta}(P)e^{i\pi\Theta(P)},
\end{equation}
where $\mathscr{M}$ is the open string amplitude without the
coupling constant $g$, and we sum over all the orderings $P$ of the
open strings.

If there are open strings on the boundary of $D_2$, we can insert
the open string vertices into the amplitude \eqref{AD_2N1}. Because
\eqref{AD_2N1} is already an amplitude for open strings on the real
axis except for a phase factor, we just increase the number of the
open strings on the boundary of $D_2$ and adjust the phase factor to
make the integrals in the right branch cuts. The phase factor should
be adjusted because we must consider the interactions between closed
and open strings. Then we have
\begin{equation}\label{(N, M)D2relations}
 \mathscr{A}_{D_2}^{(N,
M)}=\epsilon_{\alpha\beta\gamma}\mathscr{A}_{D_2}^{(N,
M)\alpha\beta\gamma}=\left(\frac{i}{4}\right)^{N-1}\kappa^{N-1}g^M\epsilon_{\alpha\beta\gamma}
\sum\limits_P\mathscr{M}^{(2N,M)\alpha\beta\gamma}(P)e^{i\pi\Theta'(P)},
\end{equation}
where we have defined the coordinates of the left-moving open
strings are $\xi_1,...,\xi_N$, those of right-moving open strings
are $\xi_{1+N+M},...,\xi_{2N+M}$ and the coordinates of other open
strings are $\xi_{1+N},...,\xi_{M+N}$.
\begin{equation}
\Theta'(P)=\sum\limits_{s>r}2\alpha'k'_s\cdot
k'_r\theta'(\xi_s-\xi_r),
\end{equation}
where $k'_r$ are the momentums of the open strings. If
$\xi_s>\xi_r$, $\theta'(\xi_s-\xi_r)=0$, else if $\xi_s<\xi_r$ but
$N<s,r<N+M+1$, $\theta'(\xi_s-\xi_r)=0$, otherwise
$\theta'(\xi_s-\xi_r)=1$.
 This relation can also be derived by choosing the fundamental
region as the upper half-plane, then repeat the similar steps in the
case of $N$ closed strings on $D_2$. We can see if $M=0$,
\eqref{A_(D_2)^(N,M)} gives the relation for $N$ closed strings on
$D_2$\eqref{D2relations} and if $N=0$ it gives the open string tree
amplitude\eqref{open string amplitude}.

 By comparing the relations \eqref{D2relations} with KLT
factorization relations \eqref{KLT relations}, we can see, in
\eqref{KLT relations}, the left- and the right-moving sectors are
independent of each other. In \eqref{D2relations}, they are not
independent of each other. The interactions connect the two open
string amplitudes into a single one. because the interactions
between the two sectors are just the open string interactions, the
amplitudes for $N$ closed strings then can be given by tree
amplitudes for $2N$ open strings.

We can consider the relations on $D_2$ as any closed strings can be
splitted into two open strings. Each open string catch half of the
momentum of the closed string. Move the open strings corresponding
to the two sectors of closed strings onto the boundary of $D_2$.
Then an amplitude for $N$ closed strings and $2M$ open strings on
$D_2$ is given by an amplitude for $N+2M$ open strings.

In  \eqref{(N, M)D2relations}, the $D_2$ amplitudes have been given
by the a sum of open string partial amplitudes with $2N+M$ external
legs correspondingly. We have to sum over all the orderings of the
open strings in this relation. However, as in the case of
$S_2$\cite{9}, the contour treatment\cite{29} can reduce the number
of the terms in this relation. The main points of the treatment
of\cite{29} is there are relations among open string partial
amplitudes. Then any open string partial amplitudes can be expressed
in terms of a minimal basis. All the $M$-point open string partial
amplitude can be expressed in terms of the minimal basis of $(M-3)!$
independent partial amplitudes. Then for the $(N, M)$ case, the
amplitude can be given by $(2N+M-3)!$ open string partial
amplitudes.

If we consider the interactions between open strings attached to a
D$p$-brane and closed strings, we should do appropriate replacements
in the right-moving sector. For example, if the external legs are
gravitons, we just need to replace the momenta
$\frac{1}{2}k^{\mu}_r$ corresponding to the right-moving sector by
$\frac{1}{2}D^{\mu}_{\nu}\cdot k^{\nu}_r$ and replace the
polarization tensor $\epsilon_{\mu\nu}$ by
$\epsilon_{\mu\lambda}D^{\lambda}_{\nu}$ in the relation\eqref{(N,
M)D2relations}\cite{25, 26, 27}. Where $D^{\mu}_{\nu}$ is defined as

\begin{equation}
\left(
\begin{array}{cccccc}
 1 & \text{ } & \text{ } & \text{ } & \text{ } & \text{ }\\
 \text{ } & \ddots & \text{ } & \text{ } & \text{ } & \text{ } \\
 \text{ } & \text{ } & 1 & \text{ } & \text{ } & \text{ }\\
 \text{ } & \text{ } & \text{ } & -1 & \text{ }  & \text{ }\\
 \text{ } & \text{ } & \text{ } & \text{ } & \ddots &\text{ }\\
 \text{ } & \text{ } & \text{ } & \text{ } & \text{ } & -1
\end{array}
\right).
\end{equation}
Then this relation reveals the amplitudes between $N$ closed string
and $M$ open strings on a D$p$-brane can be given by $2N+M$-point
open string partial amplitudes. Though in \cite{25, 26, 27}, $(2,
0)$ amplitude and $(1, 2)$ amplitude are four-point open string
amplitudes upon a certain identification between the momenta and
polarizations, in general case, there is a phase factor in the
relations.

 In the low energy
limit of an open string theory, gravitons are closed string states
and gauge particles are open string states. Then in this case, the
KLT factorization relations do not hold. We should use one amplitude
for $2N$ gauge particles instead of the product of two amplitudes
for $N$ gauge particles to give an amplitude for $N$ gravitons.

\section{Relations between amplitudes on $RP_2$ and open
string tree amplitudes} \label{relations on RP_2} In this section,
we will explore the amplitudes on $RP_2$\cite{1, 2, 28}. We first
show the correlation functions on $RP_2$ can not be factorized by
the left- and the right-moving sectors. The two sectors are
connected together. Then we will give the relations between
amplitude on $RP_2$ and tree amplitude for open strings.

$RP_2$ is an unoriented surface, it can be derived by identifying
the diametrically opposite points on $S_2$. It can be considered
as a sphere with a crosscap. With this equivalence, the waves must
be reflected at the crosscap. the reflection waves of the
left-moving waves are in the right-moving sector and the
reflection waves of the right-moving waves are in the left-moving
sector. The waves must interact with their reflection waves, thus
the two sectors must interact with each other. This is similar
with the case of $D_2$.
\newline
\begin{figure}[tbp]
\begin{center}
\includegraphics[width=0.5\textwidth]{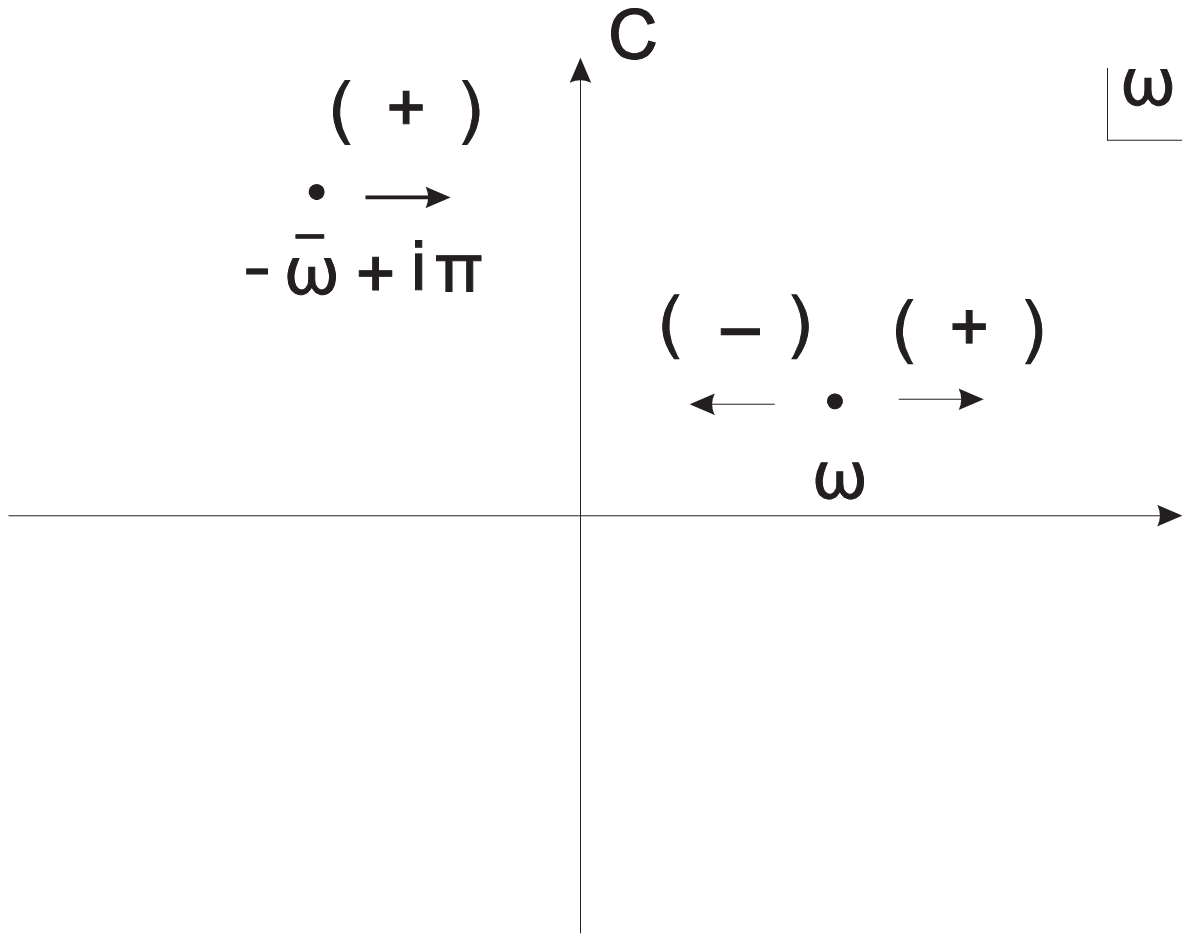}
\end{center}
\caption{Only the annihilation modes are reflected at the crosscap
of $RP_2$.} \label{fig3}
\end{figure}
\newline
Particularly, the correlation function on $RP_2$ is given as
\begin{equation}\label{RP2 correlation}
\left<0\mid\mathscr V_N(\omega,
\tilde{\omega})...\mathscr{V}_1(\omega, \tilde{\omega})\mid
C\right>,
\end{equation}
where $\mid C\rangle=C\mid0\rangle$ is the boundary sate for
$RP_2$\cite{20, 21}. The bosonized boundary operator $C$ is
\begin{equation}
C=\exp{(\sum\limits_{n=1}^\infty
(-1)^na_n^{\dagger}\cdot\tilde{a}_{n}^{\dagger})}\left|0\right>_{X}
\otimes\exp{(\sum\limits_{n=1}^\infty
(-1)^nb_n^{\dagger}\circ\tilde{b}_{n}^{\dagger})}\left|0\right>_{\phi}
\otimes\exp{(\sum\limits_{n=1}^\infty
(-1)^nc_n^{\dagger}\tilde{c}_{n}^{\dagger})}\left|0\right>_{\phi_6}.
\end{equation}
In this case we can see, the image point of $\omega$ is
$-\bar{\omega}+i\pi$. When we move $C$ to the left of a vertex
operator it commute with the creation modes and the zero modes of
the vertex operator. It does not commute with the annihilation
modes $\mathscr{V}^{(-)}_L(\omega)$ and
$\tilde{\mathscr{V}^{(-)}_R}(\bar{\omega})$. This means only the
annihilation modes can be reflected at the crosscap(see
fig\ref{fig3}). After moving the boundary operator to the left of
the annihilation modes $\mathscr{V}^{(-)}_L(\omega)$ and
$\tilde{\mathscr{V}^{(-)}_R}(\bar{\omega})$, the images
$\tilde{\mathscr{V}}^{(+)}_L(-\omega-i\pi)$ and
$\mathscr{V}^{(+)}_R(-\bar{\omega}+i\pi)$ are created
respectively. We move the boundary operator to the left of all the
vertex operators. Then use the creation operators in the boundary
operator to annihilate the state $\langle0\mid$. The correlation
becomes
\begin{equation}\label{correlation function on RP2}
\begin{split}
&\left<\mathscr{V}^{(+)}_L(\omega_N)\mathscr{V}^{(-)}_L(\omega_N)\mathscr{V}^{(+)}_R(-\bar{\omega}_N+i\pi)...\mathscr{V}^{(+)}_L(\omega_1)\mathscr{V}^{(-)}_L(\omega_1)\mathscr{V}^{(+)}_R(-\bar{\omega}_1+i\pi)\right>
\\\times &\left<\tilde{\mathscr{V}}^{(+)}_R(\bar{\omega}_N)\tilde{\mathscr{V}}^{(-)}_R(\bar{\omega}_N)\tilde{\mathscr{V}}^{(+)}_L(-\omega_N-i\pi)...\tilde{\mathscr{V}}^{(+)}_R(\bar{\omega}_1)\tilde{\mathscr{V}}^{(-)}_R(\bar{\omega}_1)\tilde{\mathscr{V}}^{(+)}_L(-\omega_1-i\pi)\right>
\\\times &\left<\mathscr{V}_0(\omega_N, \tilde{\omega}_N)...\mathscr{V}_0(\omega_1,
\tilde{\omega}_1)\right>.
\end{split}
\end{equation}
As in the case of $D_2$, the first correlation function in
\eqref{correlation function on RP2} only contain $a$, $b$, $c$ and
$a^{\dag}$, $b^{\dag}$, $c^{\dag}$. When we move the left-moving
modes of a vertex operator $\mathscr{V}^{(-)}_L(\omega_r)$ to the
right of the operator $\mathscr{V}^{(+)}_L(\omega_s)$, we get the
interaction in the left-moving sector. When we move
$\mathscr{V}^{(-)}_L(\omega_r)$ to the right of
$\mathscr{V}^{(+)}_R(-\bar{\omega}_s+i\pi)$, we get the
interaction between the left- and the right-moving sectors. In the
same way, the second correlation function in \eqref{correlation
function on RP2} gives the interactions in the right-moving sector
and those between the two sectors. Thus the correlation function
can not be factorized by the two sectors. Interactions connect the
two sectors together.

Now we consider the amplitude for $N$ closed strings on $RP_2$. We
calculate the correlation function, integral over the fundamental
region and divide the integrals by the volume of the CKG\cite{1,
2, 22, 23} on $RP_2$. As we have done in the case of $D_2$, we
also extend the integral region to the complex pane, rotate the
$y$ integrals to the real axis and redefine the integral
variables. The amplitude for $N$ closed strings on $RP_2$ can be
given as
\begin{align}\label{ARP2superstring}
\nonumber\mathscr{A}_{RP_2}^{(N)}&=\kappa^{N-1}\left(\frac{1}{2}\right)^{N-1}\int
\prod\limits_{i=1}^Nd\xi_id\eta_i\frac{|1+\xi_o\eta_o|^2}{2\pi
d\xi_o\eta_o}
\\&\nonumber\times\prod\limits_{s>r}(\xi_s-\xi_r)^{\frac{\alpha'}{2}k_r\cdot
k_s+\lambda_r\circ\lambda_s-q_rq_s}(\eta_r-\eta_s)^{\frac{\alpha'}{2}k_r\cdot
k_s+\tilde{\lambda}_r\circ\tilde{\lambda}_s-\tilde{q}_r\tilde{q}_s}
\prod\limits_{r,
s}(1+(\xi_r\eta_s)^{-1})^{\frac{\alpha'}{2}k_r\cdot
k_s+\lambda_r\circ\tilde{\lambda}_s-q_r\tilde{q}_s}
\\&\nonumber\times \exp{\sum\limits_{r=1}^N\left(\sum\limits_{i=1}^{n_r}\sum\limits_{j=1}^{\tilde{n}_s}\left(-\frac{\alpha'}{2}\right)\epsilon_r^{i}\cdot\bar{\epsilon}_r^{j}
-\sum\limits_{i=1}^{m_r}\sum\limits_{j=1}^{\tilde{m}_s}\varepsilon_r^{i}\circ\bar{\varepsilon}_r^{j}\right)(1+\xi_r\eta_r)^{-2}}
\\&\times \exp{\sum\limits_{s>r}\left[\left(\sum\limits_{i=1}^{n_r}\sum\limits_{j=1}^{n_s}\left(-\frac{\alpha'}{2}\right)\bar{\epsilon}_r^{(i)}\cdot\epsilon_s^{(j)}
-\sum\limits_{i=1}^{\tilde{m}_r}\sum\limits_{j=1}^{m_s}\bar{\varepsilon}_r^{(i)}\circ\varepsilon_s^{(j)}\right)(1+\eta_r\xi_s)^{-2}+c.c.\right]}
\\&\nonumber\times\exp{\left[-\sum\limits_{s>r}\left(\sum\limits_{i=1}^{n_r}\sum\limits_{j=1}^{n_s}\left(-\frac{\alpha'}{2}\right)\epsilon_r^{(i)}\cdot\epsilon_s^{(j)}-\sum\limits_{i=1}^{m_r}\sum\limits_{j=1}^{m_s}\varepsilon_r^{(i)}\circ\varepsilon_s^{(j)}\right)(\xi_s-\xi_r)^{-2}+c.c.
\right]}
\\&\nonumber\times \exp{\sum\limits_{r\neq s}\left[\left(\sum\limits_{i=1}^{n_s}\left(-\frac{\alpha'}{2}\right)k_r\cdot\epsilon_s^{(i)}
-\sum\limits_{i=1}^{m_s}\lambda_r\circ\varepsilon_s^{(i)}\right)((\xi_r-\xi_s)^{-1}+(-\eta_r^{-1}-\xi_s)^{-1})+c.c.\right]}
\\&\nonumber\times
\exp{\sum\limits_{r=1}^N\left[\left(\left(-\frac{\alpha'}{2}\right)k_r\cdot\sum\limits_{i=1}^{n_r}\epsilon_r^{(i)}
-\lambda_r\circ\sum\limits_{i=1}^{m_r}\varepsilon_r^{(i)}\right)((-\eta_r^{-1}-\xi_r)^{-1}+{\xi_r}^{-1})+c.c.\right]}|_{multilinear}.
\end{align}
Then the amplitude has been given by real integrals. The
interactions in one sector are the open string interactions. The
interaction between left- and right-moving sectors can be
considered as interactions between open strings inserted at
$\xi_r$ and $(-\eta_s)^{-1}$. $\frac{1}{-\bar{\eta_s}}$ can be
considered as a time reverse and a twist in the right-moving
sector. Then the interactions between left- and right-moving
sectors can be regarded as interactions between left- and
right-moving open strings with a time reverse and a twist in the
right-moving sector(see fig\ref{fig1}.(c)).

From fig. \ref{fig1}.(c), we can see, if we twist the right-moving
sector and reverse the time in the right-moving sector, we will get
an open string tree amplitude. In fact, we can replace all the
$\eta_r$ by $-\frac{1}{\eta_r}$.  Then by using the mass-shell
condition, the interactions between the two different sectors as
well as in one sector become the interactions between open strings.
Redefine the variables in the right-moving sector by
Eq.\eqref{variable redefine}.
 The amplitude on $RP_2$ then becomes
\begin{equation}\label{ARP_2N1}
\begin{split}
\mathscr{A}_{RP_2}^{(N)}&=-{\left(\frac{i}{4}\right)}^{N-1}\kappa^{N-1}\int
\prod\limits_{i=1}^{2N}d\xi_i\frac{|\xi_a-\xi_b||\xi_b-\xi_c||\xi_c-\xi_a|}{d\xi_ad\xi_bd\xi_c}\prod\limits_{s>r}|\xi_s-\xi_r|^{\frac{\alpha'}{2}k_r\cdot
k_s}(\xi_s-\xi_r)^{\lambda_r\circ\lambda_s-q_rq_s}
\\&\times\exp{\left[\sum\limits_{s>r}\left(\sum\limits_{i=1}^{n_r}\sum\limits_{j=1}^{n_s}\left(2\alpha'\right)\epsilon_r^{(i)}\cdot\epsilon_s^{(j)}+\sum\limits_{i=1}^{m_r}\sum\limits_{j=1}^{m_s}\varepsilon_r^{(i)}\circ\varepsilon_s^{(j)}\right)(\xi_s-\xi_r)^{-2}
\right]}
\\&\times \exp{\left[\sum\limits_{r\neq s}\left(\sum\limits_{i=1}^{n_s}\left(-2\alpha'\right)k_r\cdot\epsilon_s^{(i)}
-\sum\limits_{i=1}^{m_s}\lambda_r\circ\varepsilon_s^{(i)}\right)(\xi_r-\xi_s)^{-1}\right]}|_{multilinear}e^{i\pi\Theta(P)}
.
\end{split}
\end{equation}
This amplitude is different from $D_2$ amplitude by a factor $-1$.
It is caused by the difference between the measure of the CKG on
$RP_2$ and $D_2$. When we change the topology, this $-1$ appears.
The phase factor only depends on the ordering of the open strings.
We can break the integrals into pieces as in the case of $D_2$ and
keep the multilinear terms of the polarization tensers. Then
Eq.\eqref{ARP_2N1} becomes
\begin{equation}\label{RP2relations}
\mathscr{A}_{RP_2}^{(N)}=\epsilon_{\alpha\beta}\mathscr{A}_{RP_2}^{(N)\alpha\beta}=-\left(\frac{i}{4}\right)^{N-1}\kappa^{N-1}\epsilon_{\alpha\beta}\sum\limits_P\mathscr{M}^{(2N)\alpha\beta}(P)e^{i\pi\Theta(P)}.
\end{equation}

As in the case of $D_2$, KLT factorization relations\eqref{KLT
relations} do not hold on $RP_2$. The left- and the right-moving
sectors are not independent of each other again. The interactions
between the two sectors connect them into a single sector. Since the
interactions between the two sectors are just the those between open
strings, the two open string tree amplitude in the case of $S_2$ are
connected into one amplitude for open strings. In the
relation\eqref{RP2relations}, we also sum over all the orderings of
the external legs of the open strings. By using the same method in
\cite{29}, the relations on $RP_2$ for $N$ closed strings can be
reduced to $(2N-3)!$ terms.

From the relations \eqref{D2relations} and \eqref{RP2relations} we
can see, the amplitudes on $D_2$ and $RP_2$ with same external
closed string states are equal except for a factor $-1$. In fact,
after we transform the complex variables into real ones, the image
of a point $\xi_r$ in the left-moving sector becomes
$\frac{1}{\eta_r}$ on $D_2$ and $-\frac{1}{\eta_r}$ on $RP_2$. The
minus means a twist in the right-moving sector. After this twist,
the amplitude on $RP_2$ becomes that on $D_2$ except for a factor
$-1$. Then if we consider a theory containing both $D_2$ and $RP_2$,
the amplitudes with same external states cancel out. However, if we
consider T-duality, the interactions on $D_2$ becomes interactions
between closed strings and D-brane, while the interactions on $RP_2$
becomes interactions between closed strings and O-plane. As we have
seen in section \ref{relations on D_2}, we should make appropriate
replacement on the momenta and polarizations in the right-moving
sector to give the relations in $D_2$ case. Under the T-duality, we
also need to replace the vertex operators on $RP_2$ by new
ones\cite{30}. We take the massless NS-NS vertex operator as an
example. The vertex operator after T-duality becomes
\begin{equation}\label{RP_2 vertex}
\begin{split}
&\mathscr{V}^{RP_2}(\epsilon, k, z, \bar{z}) \\&=\frac{1}{2}\left(
\epsilon_{\mu\nu}:\mathscr{V}^\mu_{\alpha}(k,
z)::\tilde{\mathscr{V}}^{\mu}_{\beta}(k,
\bar{z}):+(D\cdot\epsilon^T\cdot
D)_{\mu\nu}:\mathscr{V}^{\mu}_{\alpha}(k\cdot D,
z)::\tilde{\mathscr{V}}^{\nu}_{\beta}(k\cdot D, \bar{z}):\right).
\end{split}
\end{equation}
Here $\mathscr{V}^{\mu}_{\alpha}(p, \epsilon)$ in $0$ and $-1$
picture are
\begin{equation}
\begin{split}
\mathscr{V}^{\mu}_{-1}(k, z)&=e^{-\phi(z)}\psi^{\mu}(z)e^{ik\cdot
X(z)}
\\\mathscr{V}^{\mu}_0(k, z)&=(\partial X^{\mu}(z)+ik\cdot
\psi(z)\psi^{\mu}(z))e^{ik\cdot X}.
\end{split}
\end{equation}
The second term in \eqref{RP_2 vertex} can also be given by
replacing $\epsilon_{\mu\nu}$ and $k^{\mu}$ in the original vertex
by $(D\cdot\epsilon^T\cdot D)_{\mu\nu}$ and $(k\cdot D)^{\mu}$
respectively. Now we consider the amplitudes for $N$ NS-NS strings.
Each vertex operator\eqref{RP_2 vertex} have two terms, each term
can be considered as a vertex operator on $RP_2$ under appropriate
replacement. The amplitude then is given by $2^N$ terms, each term
can be obtained from the amplitude before T-duality by appropriate
replacements. Then each term can be given by  partial amplitudes for
$2N$ open strings again. So the $RP_2$ relation gives the amplitudes
for closed strings scattering from an O-plane by open string
amplitudes. Generally, under the T-duality, the $D_2$ amplitudes can
not be canceled by the $RP_2$ amplitudes\cite{30}.

In the low energy limit of an unoriented string theory, the
amplitudes for closed strings on $RP_2$ contribute to the amplitudes
for gravitons. Then the KLT factorization relations do not hold in
this case as in the case of $D_2$. The amplitudes for $N$ gravitons
can not be factorized by two amplitudes for $N$ gauge particles.
They can be given by an amplitude for $2N$ gauge particles.

\section{Conclusion}
\label{conclusion} In this paper, we investigated the relations
between closed and open strings on $D_2$ and $RP_2$. We have shown
that the KLT factorization relations do not hold for these two
topologies. The closed string amplitudes can not be factorized by
tree amplitudes for left- and right-moving open strings. However,
the two sectors are connected into a single sector. We can give the
amplitudes with closed strings in these two cases by amplitudes in
this single sector. The terms in the relations on $D_2$ and $RP_2$
can be reduced by contour deformations.

Under the T-duality, the relations on $D_2$ and $RP_2$ give the
amplitudes between closed strings scattering from D-brane and
O-plane respectively by open string partial amplitudes.

In the low energy limits of these two cases, we can not use KLT
relations to factorize amplitudes for gravitons into products of two
amplitudes for gauge particles. Interactions between the ``left-''
and the ``right-'' moving gauge fields connect the two amplitudes
into one. Then an graviton amplitude in these two cases can be given
by one amplitude for both left- and right-moving gauge particles.

 The relations for other topologies have not
been given. However, we expect there are also some relations between
closed and open string amplitudes. If there are more boundaries and
crosscaps on the world-sheet, the boundaries and the crosscaps also
connect left- and the right-moving sectors, then in these cases, KLT
factorization relations do not hold.

\section*{Acknowledgement}
We would like to thank C.Cao, J.L.Li, Y.Q.Wang and Y.Xiao for
helpful discussions. We would like to thank the referee for many
helpful suggestions. The work is supported in part by the NNSF of
China Grant No. 90503009, No. 10775116, and 973 Program Grant No.
2005CB724508.

\end{document}